\documentclass[%
 reprint,
 amsmath,amssymb,
 aps,
]{revtex4-1}

\usepackage{graphicx}
\usepackage{dcolumn}
\usepackage{bm}

\usepackage{tikz}
\usepackage{pgfplots}

\newtheorem{theoremm}{Theorem}

\begin{document}

\title{The Clouds in Asynchronous Cellular Automata}

\author{Souvik Roy}
\email{svkr89@gmail.com}
\altaffiliation[]{Department of Information Technology, Indian Institute of Engineering Science and Technology, Shibpur, Howrah, West Bengal, India 711103.}
\author{Sukanta Das}
\email{sukanta@it.iiests.ac.in}
\altaffiliation[]{Department of Information Technology, Indian Institute of Engineering Science and Technology, Shibpur, Howrah, West Bengal, India 711103.}

\date{\today}


\begin{abstract}
This article introduces the notion of clouds in asynchronous cellular automata (ACAs). We show that the cloud behaviour of ACAs has similarity with natural clouds across the sky, election model of parliamentary democratic system, and electron cloud around nucleus. These systems, therefore, can be modelled by the ACAs.  
\end{abstract}

\pacs{Valid PACS appear here}
\maketitle


When we look up at the sky, we see the clouds sailing across the sky. From the global view, the clouds are associated with multiple attractor basins. It is often unknown where a cloud can find her ultimate attractor basin and create a downpour of rain on the earth. In this article, we explore natural cloud phenomenon, which is observed in simple asynchronous cellular automata (CAs) models. In such CAs, the cells are considered independent to some extent. A traditional (synchronous) convergent cellular automaton (CA) with an initial configuration always converges to a specific attractor following a specific deterministic path. On the contrary, a CA with same initial configuration may approach to different attractor basins for different runs when the cells of the CA are updated independently. It may also be possible in such a system that for an initial configuration, the system always converges to a specific point attractor. Obviously, the basin of attraction in the convergence-dynamics of a system with only one attractor basin includes all the configurations for which the system approaches to the only attractor. However, the dynamics of the system under consideration differs from the traditional deterministic one. Under this system, the travelling path towards a single attractor basin may change in different runs. In this perspective, this independent system, itself, breaks the traditional deterministic concept of basins of attraction in CA \cite{Wue92,Hanson1992,ITO1988318}.

\begin{figure}[ht]\centering 
\includegraphics[width=2.5in]{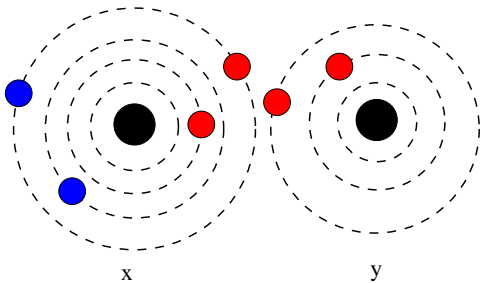} 
\caption{Graphical visualization of clouds in CA.}
\label{orbit}
\end{figure}

Now, we can define the notion of clouds for an asynchronous CA (ACA) system \cite{PhysRevE.59.3876,PhysRevE.91.042110,SCHONFISCH1999123,jca/Fates14}. One can assume that an attractor is associated with a number of orbits. Depending on the number of travelling time steps to reach to the corresponding attractor, an initial configuration is placed in an orbit. An initial configuration may simultaneously exist at multiple orbits for a single attractor, because different time steps may be needed in different runs to converge to the attractor. It may also be possible that an initial configuration can simultaneously be at multiple orbits for multiple attractors when the system converges to different attractors from the same initial configuration.  In Fig.~\ref{orbit}, blue initial configuration is simultaneously present at third and fourth orbits of $x$ (in black) attractor. Presence of a configuration in orbit $i$ indicates that the CA takes $i$ time steps to reach to the attractor from the configuration. The red initial configuration in Fig.~\ref{orbit} exists at multiple orbits for both the $x$ and $y$ attractors. Note that, an orbit can be associated with multiple initial configurations. Therefore, the configurations can be seen as clouds around the attractors. This notion of clouds in the asynchronous CA reminds us the concept of electron cloud.

\begin{figure*}[ht]\centering 
\includegraphics[width=6.5in]{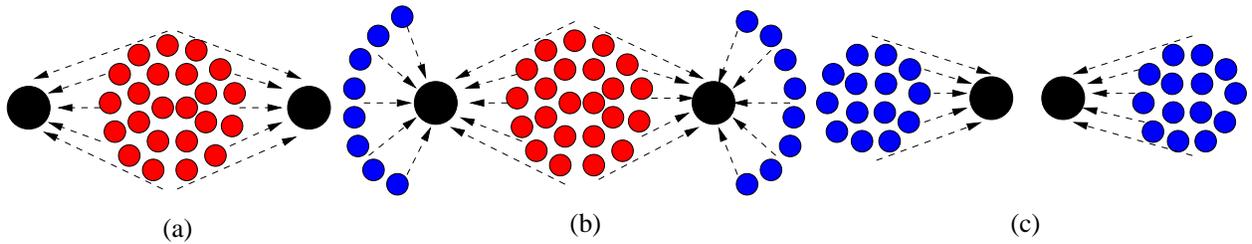} 
\caption{Two attractor (in black) (a) eccentric; (b) partially eccentric; and (c) deterministic cloud system where red initial configurations are confused configuration and blur configurations are associated with deterministic destination.}
\label{abst}
\end{figure*}

To sum up, an initial configuration for which the CA converges to a specific attractor for every run or different attractors for different runs, the concept of cloud is applicable to that initial configuration. However, convergence at different attractors for different runs injects an extra flavour in the system. We name an initial configuration, for which the CA converges to different point attractors for different runs, as \textbf{confused configuration}. In Fig.~\ref{abst}, a red initial configuration depicts the graphical visualization of confused configuration which are linked to both the attractor basins (in black). In this perspective, depending on the presence of confused configuration, we classify the system into three classes - \textbf{eccentric cloud}, \textbf{partially eccentric cloud} and \textbf{deterministic cloud} system.

In an eccentric cloud system, for every possible initial configuration, the system converges to different attractors for different runs. Fig.~\ref{abst}(a) shows a two-attractor eccentric cloud system where for every initial configuration (in red), the system may converge to any one of the two point attractors (in black). The dynamics of this system reminds us the clumsy clouds in the sky which do not really have a predictable destination. Obviously, a CA with point attractor as initial configuration always converges to itself in a deterministic manner.

In a partially eccentric cloud system, some initial configurations are confused configuration and some are not. In a two-attractor system of Fig.~\ref{abst}(b), red initial configuration is a confused configuration and blue configuration is with fixed destination. Remark that, partially eccentric cloud system may catch the dynamics of election model of a parliamentary democratic system. In an election, there are {\em certain} or {\em lock} votes who are solidly behind or partisan to a particular candidate and will not consider changing their minds whatever the opposition says. However, there also exist some {\em swing} voters whose votes are unpredictable. Here, swing voters can be modeled by confused configurations. In American politics, many centrists, liberal Republicans and conservative Democrats are considered `swing voters' since their voting patterns can not be predicted with certainty. 

Every initial configuration is always linked to a specific point attractor in the deterministic cloud system. In Fig.~\ref{abst}(c), every blue initial configuration is associated with a deterministic destination attractor. One can imagine this type of system as electron cloud model where we cannot know exactly where an electron is at any given time, but the electrons are more likely to be in specific areas.

Here, we work in the rule space of binary three neighbourhood 1-D CA (known as elementary cellular automata (ECA)) under periodic boundary condition, where the cells are arranged as a ring. The next state of each CA cell is defined as $S_i^{t+1}$ = $f$($S_{i-1}^{t}$,$S_i^{t}$,$S_{i+1}^{t}$), where $f$ is the local rule and $S_{i-1}^{t}$,$S_i^{t}$ and $S_{i+1}^{t}$ are the present states of left, self and right neighbours respectively. A collection of (local) states of cells at time $t$ is referred to as a configuration of the CA. The local rule $f$: \{$0,1$\}$^3$ $\mapsto$ \{$0,1$\} can be expressed as a look-up table (see Table~\ref{Trules}). We call each argument of $f$ as Rule Min Term (RMT) \cite{SethiRD16}. An RMT ($x,y,z$) is generally represented by its decimal from $r$ = $4*x + 2*y + z$ (row $2$ of Table~\ref{Trules}). However, the decimal counterpart of the eight next state is referred as ``rule'' \cite{compref351}. There are $2^8$($256$) ECA rules, out of which $88$ are minimal representative rules and the rests are their equivalents \cite{Li90thestructure}. The CAs are updated asynchronously where a single cell is selected uniformly at random for update. This update scheme also implies a scheme where all but two neighbouring cells may act simultaneously. Such CAs are referred as {\em fully} asynchronous CA (ACA) \cite{SethiRD16}.

\begin{table}   
\centering
\scriptsize
\begin{tabular}{cccccccccc} \hline  
($x,y,z$)   &  111 & 110 & 101 & 100 & 011 &  010 &  001 &  000 &  \\    
(RMT) & (7) & (6) & (5) & (4) & (3) & (2) & (1) & (0) &  Rule\\ \hline 
   &  0  &  1 &  1  &  0  &   1  &  0  &   0  &   0   & 104\\ 
f($x,y,z$)   &  1  &  0 &  1  &  0  &   1  &  0  &   0  &   0   & 168\\  
   &  1  &  0 &  1  &  0  &   1  &  0  &   1  &   0   & 170\\  \hline
\end{tabular}
\caption{Look-up table for ECA 104,~168 and 170.\label{Trules}}  
\end{table}

An RMT $r$ of a rule $R$ is active if a CA cell flips its state ($1$ to $0$ or $0$ to $1$) on $r$; otherwise, the RMT $r$ is passive. An ACA configuration is a point attractor if the RMT corresponding to any three consecutive bits of the configuration is passive. That is, if an ACA reaches a point attractor, the ACA remains in that particular configuration forever. In the CAs under consideration, we have identified six special configurations which we call homogeneous configurations - \textbf{0},~\textbf{1},~\textbf{01},~\textbf{001},~\textbf{011} and \textbf{0011} \cite{130519}. In \textbf{0} and \textbf{1}, all cells are in state $0$ and $1$ respectively. Similarly, in \textbf{01}, cells are in states alternate $0$ and $1$. And, so on. The corresponding RMT sets of homogeneous configurations are referred as primary RMT sets - \{$0$\}, \{$7$\}, \{$2$,$5$\}, \{$4$,$1$,$2$\}, \{$5$,$3$,$6$\} and \{$4$,$1$,$3$,$6$\}. An ACA configuration is formed using the RMTs of one or more of these six sets. We have identified that $50$ minimal representative ACAs, out of $88$, converge to point attractor \cite{SethiRD16}. Out of this $50$ convergent ACAs, $18$ ACAs with only one point attractor always belong to deterministic cloud ACA. Now, the target is to explore the convergence dynamics of rest $32$ ACAs with multiple point attractors (see Table~\ref{t32}).

\begin{table}[h]  
\centering
\scriptsize
\begin{tabular}{|ccccccccccc|} \hline  
4&5&12&13&36&44&72&76&77&78&94\\
104&\textbf{128}&\textbf{130}&132&\textbf{136}&\textbf{138}&140&\textbf{146}&\textbf{152}&\textbf{154}&\textbf{160}\\
\textbf{162}&164&\textbf{168}&\textbf{170}&172&\textbf{178}&\textbf{184}&200&204&232&\\
 \hline
\end{tabular} 
\caption{List of 32 ACAs with multiple point attractors.} 
\label{t32} 
\end{table}

\begin{figure*}[t] 
\centering  
\begin{tabular}{@{\hspace{-0.0in}}c@{\hspace{0.1in}}c@{\hspace{-0.2in}}c}     
\includegraphics[scale=0.09]{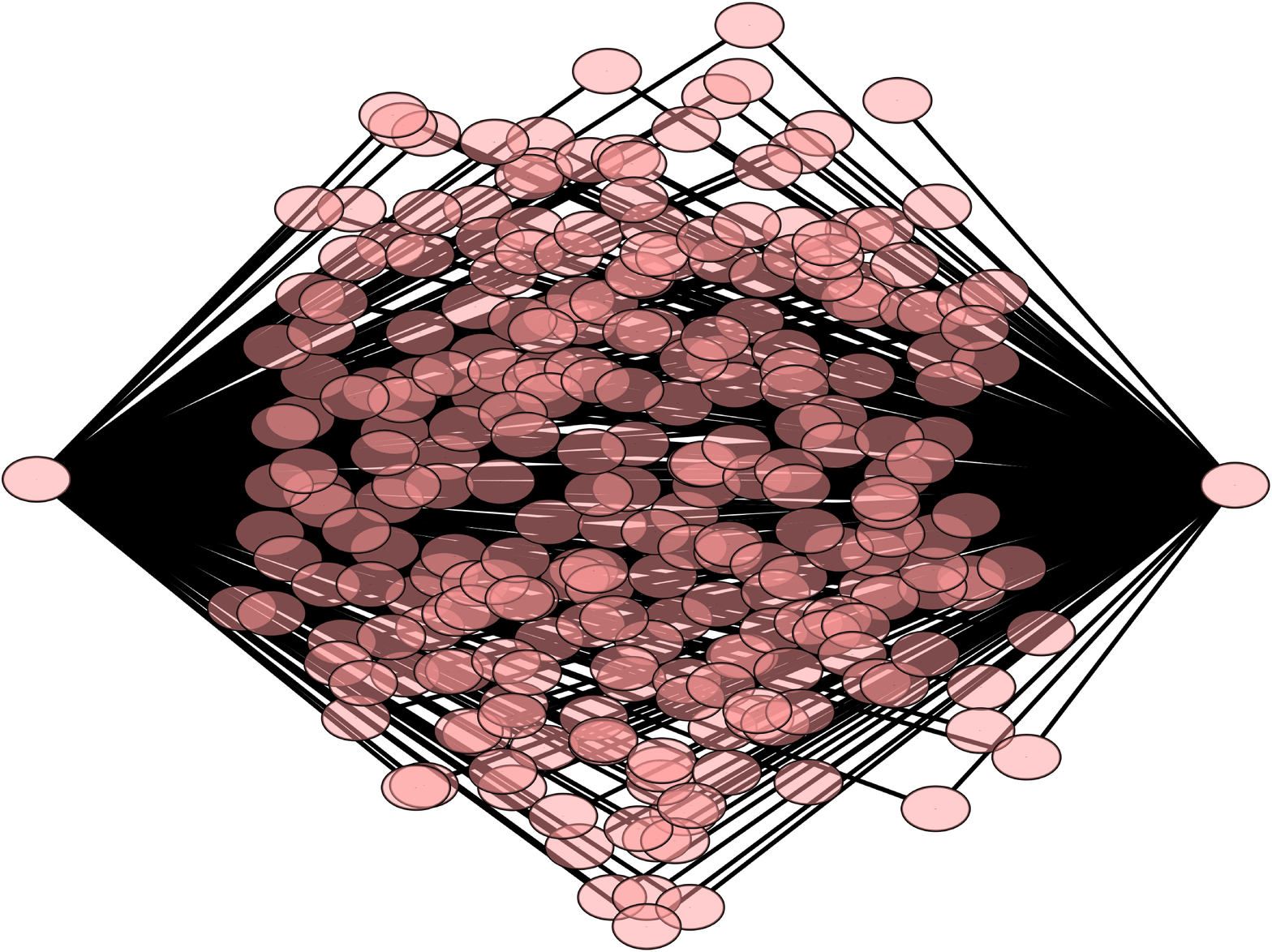} & \includegraphics[scale=0.09]{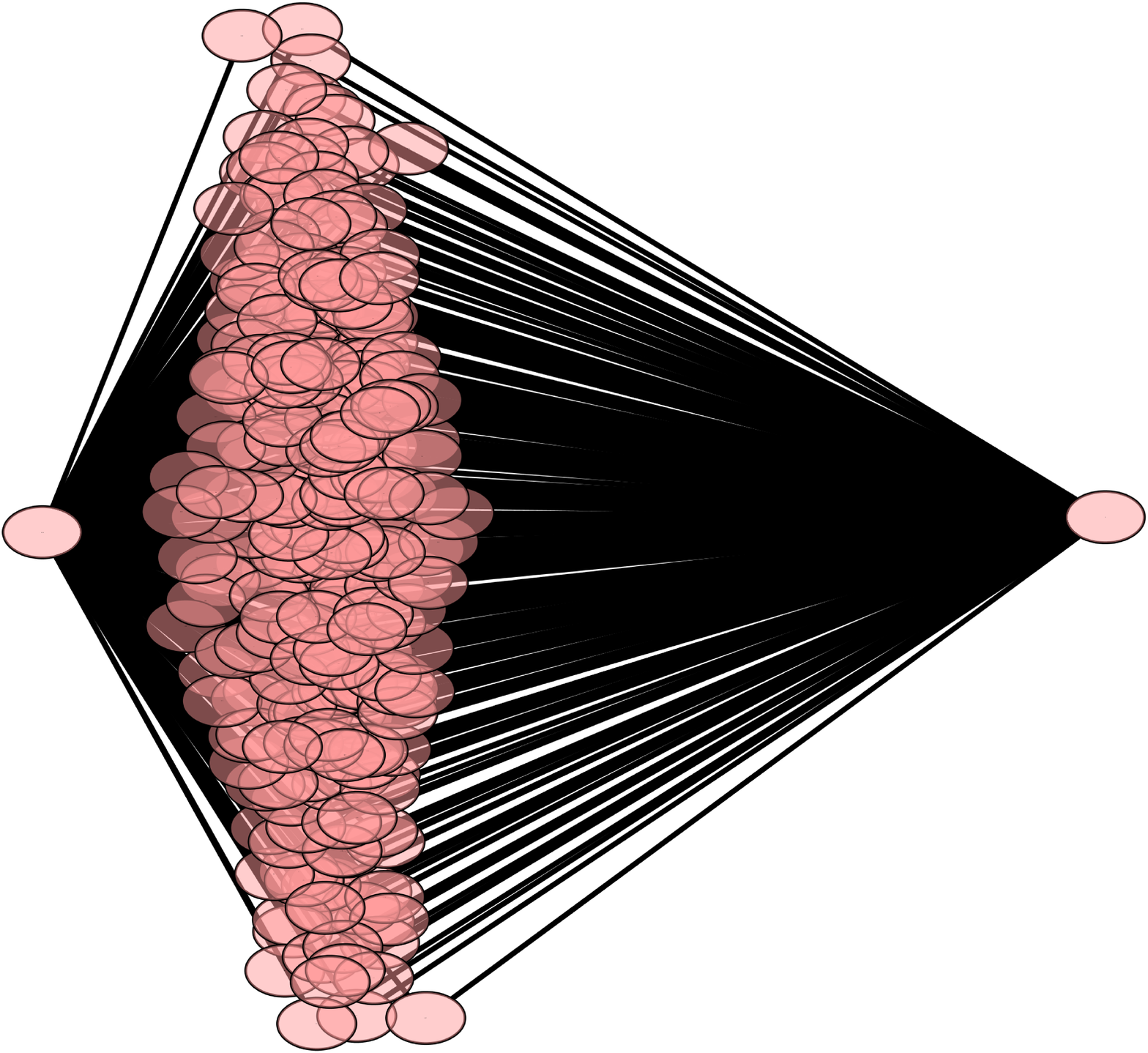}\\
(a)& (b)\\
\end{tabular} 
\caption{Convergence dynamics of $8$-cell two attractor (a) unbiased system (ACA 170); and (b) biased system (ACA 162). The left (resp. right) point attractor is referred as \textbf{0} (resp. \textbf{1}).}  
\label{twocloud1}  
\end{figure*} 

Let us start the discussion with the dynamics of two-attractor system where the point attractors are \textbf{0} and \textbf{1} (see Table~\ref{t32}, in bold). Remark that, in a two-attractor system, active RMTs $1$ and $4$ (resp. $3$ and $6$) are responsible for the travelling path towards point attractor \textbf{1} (resp. \textbf{0}). That is, active RMTs $1$ and $4$ are answerable for increasing number of $1$'s. For example, consider a configuration $001100$ with sequence of RMTs $\langle 0\textbf{1}36\textbf{4}0 \rangle$ for which the possible transitions are $001100$ $\rightarrow$ $011100 / 001110$. Whereas active RMTs $3$ and $6$ are answerable for decreasing number of $1$'s. For example configuration, the possible transitions are $110011$ $\langle 7\textbf{6}41\textbf{3}7 \rangle$ $\rightarrow$ $100011 / 110001$. Therefore, participation from both the sets \{$1,4$\} and \{$3,6$\} in the active RMT set identifies two attractor eccentric cloud system. Here, we have identified ACAs \textbf{162},~\textbf{170},~\textbf{178},~\textbf{184} as eccentric cloud systems. Let us now explore the dynamics of two-attractor eccentric cloud ACA $170$ where active RMTs are $1$ and $6$ (see Table~\ref{Trules}). Here, the participation of RMT sets \{$1,4$\} and \{$3,6$\} is equal in the active RMT set. Therefore, the probability of increasing number of $1$'s and decreasing number of $1$'s at the next time step starting from any arbitrary initial configuration is equal. In details, the probability of convergence to point attractor \textbf{0} (resp. \textbf{1}) starting from initial configuration $x$ is $P_{170}^x$(\textbf{0}) = $a/n$ (resp. $P_{170}^x$(\textbf{1}) = $b/n$) where number of $0$'s (resp. $1$'s) in $x$ initial configuration is $a$ (resp. $b$) and $a+b$ = $n$. Therefore, $\mathbb{P}_{170}$($\textbf{0}$) $\approx$ $\mathbb{P}_{170}$($\textbf{1}$), where $\mathbb{P}_{170}$($\textbf{0}$) (resp. $\mathbb{P}_{170}$($\textbf{1}$)) is the overall probability of convergence to point attractor \textbf{0} (resp. \textbf{1}). As an evidence, let us draw the convergence dynamics of ACA $170$. In Fig.~\ref{twocloud1}, if for a confused configuration, the ACA has high probability to reach to a point attractor, the configuration is placed closer to the point attractor. Here, the cloud of confused configurations, in Fig.~\ref{twocloud1}(a), is not biased towards any specific point attractor. It may also possible that the cloud of confused configurations is biased towards a specific point attractor, see Fig.~\ref{twocloud1}(b).

\begin{figure}[ht]
\centering  
\begin{tabular}{@{\hspace{-0.0in}}c}     
\includegraphics[scale=0.1]{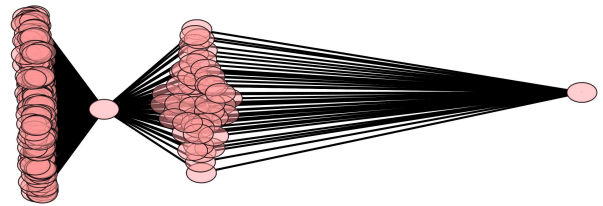} \\
\end{tabular} 
\caption{Convergence dynamics of $8$-cell two-attractor partially eccentric cloud system (ACA $168$). The left and right point attractor is referred as \textbf{0} and \textbf{1} respectively.}  
\label{twoc}  
\end{figure}

In this context, we classify the eccentric cloud systems into two classes - \textbf{unbiased} system and \textbf{biased} system. For the two-attractor system, an ACA $\delta$ is said to be unbiased system if $\mathbb{P}_{\delta}$($\textbf{0}$) $\approx$ $\mathbb{P}_{\delta}$($\textbf{1}$), whereas ACA $\delta$ is said to be biased system if $\mathbb{P}_{\delta}$($\textbf{0}$) $\gg$ $\mathbb{P}_{\delta}$($\textbf{1}$) or $\mathbb{P}_{\delta}$($\textbf{0}$) $\ll$ $\mathbb{P}_{\delta}$($\textbf{1}$). From the RMT point of view, equal participation of RMT sets \{$1$,$4$\} and \{$3$,$6$\} in the active RMT set identifies unbiased system (ACAs \textbf{170},~\textbf{178},~\textbf{184}), otherwise the system reflects the biased nature (ACA \textbf{162}). Fig.~\ref{twocloud1}(b) depicts the dynamics of biased ACA $162$ where $\mathbb{P}_{162}$($\textbf{0}$) $\gg$ $\mathbb{P}_{162}$($\textbf{1}$). 

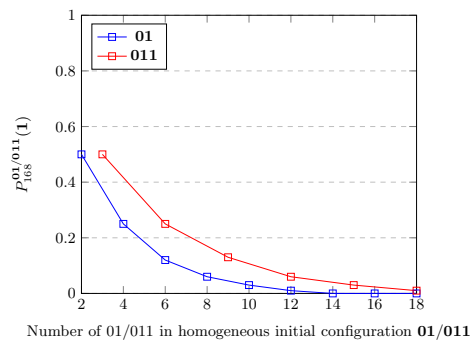
\begin{figure}
\centering 
\begin{tikzpicture}[scale=0.65]
\begin{axis}[
    title={},
   ylabel={$P_{168}^{\textbf{01}/\textbf{011}}$($\textbf{1}$)},
    xlabel={Number of 01/011 in homogeneous initial configuration \textbf{01}/\textbf{011}},
    xmin=2, xmax=18,
    ymin=0, ymax=1,
    xtick={2,4,6,8,10,12,14,16,18},
    ytick={0.00,0.20,0.40,0.60,0.80,1.00},
    legend pos=north west,
    ymajorgrids=true,
    grid style=dashed,
]
\addplot[
    color=blue,
    mark=square,
    ]
    coordinates {
    (2,0.50)(4,0.25)(6,0.12)(8,0.06)(10,0.03)(12,0.01)(14,0.00)(16,0.00)(18,0.00)
    };
    \addlegendentry{\textbf{01}}
    
\addplot[
    color=red,
    mark=square,
    ]
    coordinates {
    (3,0.50)(6,0.25)(9,0.13)(12,0.06)(15,0.03)(18,0.01)
    };
    \addlegendentry{\textbf{011}}
\end{axis}
\end{tikzpicture}
\caption{\scriptsize{ $P_{168}^{\textbf{01}/\textbf{011}}$($\textbf{1}$) as a function of number of 01/011 in the homogeneous initial configuration \textbf{01}/\textbf{011} for ACA 168.}} \label{plot}
\end{figure}

In this context, the notion of unbiased and biased nature is not applicable to the configurations leading to specific destination in a partially eccentric system. However, the notion is valid for confused configurations in the system. Remark that, participation from only one set \{$1,4$\} or \{$3,6$\} in the active RMT set identifies two-attractor partially eccentric cloud ACAs \textbf{160},~\textbf{168}. To understand the dynamics of two attractor partially eccentric cloud system, we consider ACA $168$ (see Table~\ref{Trules}). Here, it is not possible to break more than one consecutive $0$'s as RMTs $1$ and $4$ are passive. That is, a configuration with more than one consecutive $0$'s, $\cdot\cdot 110011 \cdot\cdot$, can be viewed as a sequence of RMTs $\langle \cdot\cdot 6\textbf{41}3 \cdot\cdot \rangle$. Therefore, an initial configuration with primary RMT set \{$4,1,2$\} and/or \{$4,1,3,6$\} is associated with deterministic point attractor \textbf{0}. However, the notion of unbiased or biased nature is true for a confused configuration with primary RMT sets \{$2,5$\} and/or \{$5,3,6$\} where the system can be able to converge at both the point attractors \textbf{0} and \textbf{1}. During the evolution of the confused configuration, if a single cell in state $1$ moves to state $0$, then the system converges to point attractor \textbf{0}. That is, the configuration is associated with more than one (two) consecutive $0$'s when a single cell in state $1$ moves to state $0$. As an example, one of the possible transition is $011011$ $\rightarrow$ $011001$. Recall that, this type of configurations are not breakable. Therefore, to reach point attractor \textbf{1}, the target is to update cell(s) in state $0$ at every time steps which is almost impossible for large CA size. That is, $P_{168}^{\textbf{01}/\textbf{011}}$($\textbf{1}$) $\rightarrow 0$ for large CA size. As an evidence, Fig.~\ref{plot} depicts the $P_{168}^{\textbf{01}}$($\textbf{1}$) and $P_{168}^{\textbf{011}}$($\textbf{1}$) starting from homogeneous initial configurations \textbf{01} and \textbf{011} respectively. Therefore, $\mathbb{P}_{168}$($\textbf{0}$) $\rightarrow 1$ and $\mathbb{P}_{168}$($\textbf{1}$) $\rightarrow 0$ for large CA size. To sum up, Fig.~\ref{twoc} depicts the dynamics of ACA $168$ where every confused configuration is biased towards point attractor \textbf{0} (left one).

\begin{figure}
\centering  
\begin{tabular}{@{\hspace{-0.0in}}c}     
\includegraphics[scale=0.08]{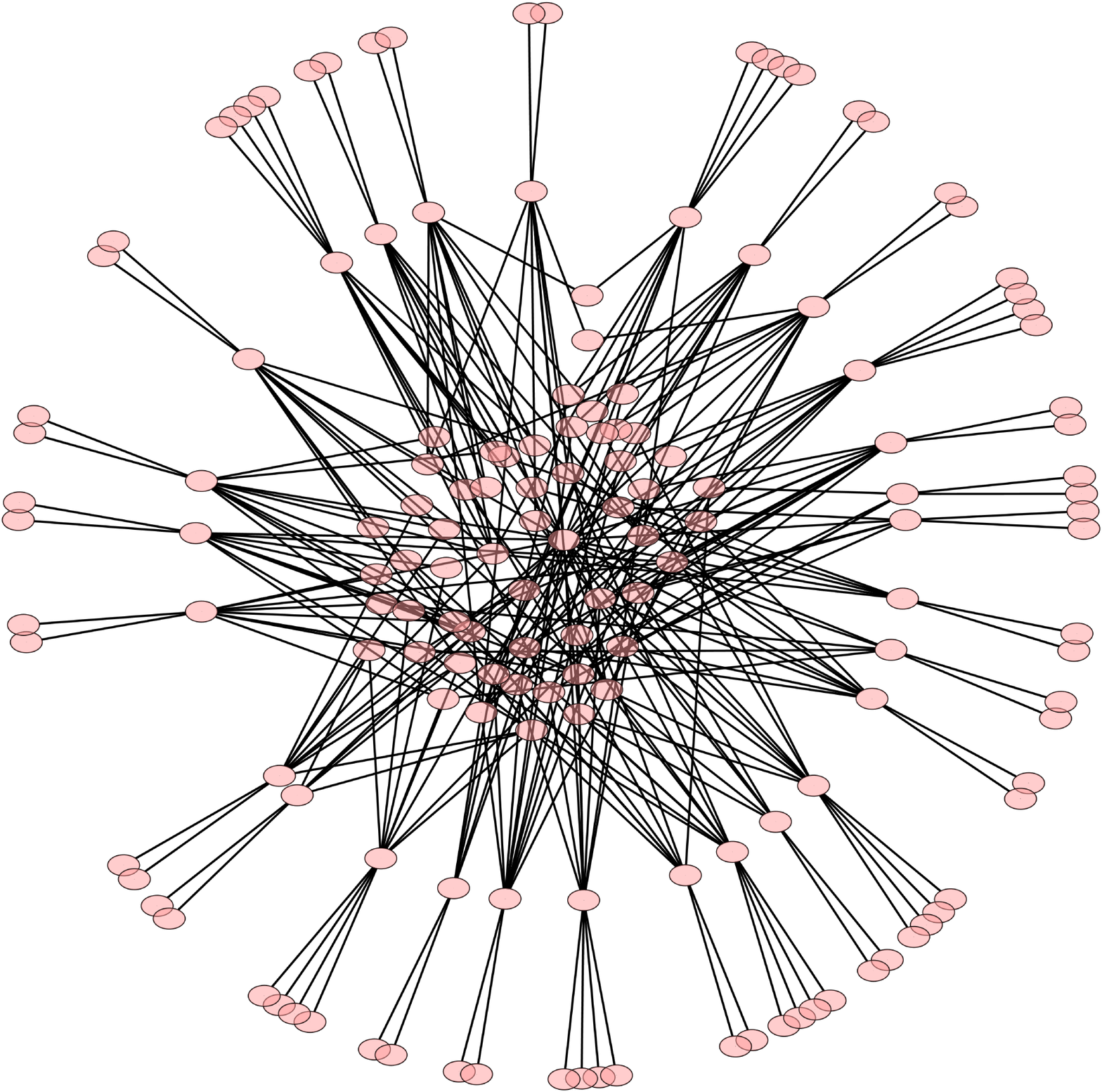} \\
\end{tabular} 
\caption{Convergence dynamics of $7$-cell multiple attractor partially eccentric cloud system (ACA $12$).}  
\label{twocloud2}  
\end{figure}

We have also identified simple two-attractor deterministic cloud system, where the system can not be able to converge at point attractor \textbf{1} for any initial configuration. Remark that, in a two-attractor system, point attractor \textbf{0} (resp. \textbf{1}) is not reachable from any initial configuration if RMT $2$ (resp. $5$) is passive. The example transitions are $00100$ $\langle 01\textbf{2}40 \rangle$ $\nrightarrow$ \textbf{0} and $11011$ $\langle 76\textbf{5}37 \rangle$ $\nrightarrow$ \textbf{1} respectively. Here, deterministic cloud ACAs \textbf{128},~\textbf{130},~\textbf{136},~\textbf{138},~\textbf{146},~\textbf{152},~\textbf{154} are associated with passive RMT $5$.

Let us now start the discussion on the convergence dynamics of multiple attractor (more than two) system with the \textbf{dual property} of RMTs. This dual property of active RMTs $0$ and $7$ indicates that these RMTs are responsible for both fixed and confused convergence journey when primary RMT sets \{2,5\},\{4,1,2\} and \{2,5\},\{5,3,6\} are responsible for point attractors respectively. As an example, RMT $0$ is accountable for fixed convergence journey $010001$ $\rightarrow$ $010101$ and confused convergence journey $000000$ $\rightarrow$ $010101 / 001001$ for ACA $77$. That is, this property identifies multiple attractor partially eccentric cloud ACAs \textbf{5},~\textbf{13},~\textbf{76},~\textbf{77},~\textbf{78},~\textbf{94}.

The observations made on the convergence of partially eccentric cloud system are probably the most interesting result of the study. Fig.~\ref{twocloud2} shows multiple attractor partially eccentric cloud system for ACA $12$ where primary RMT sets \{$0$\},\{$2,5$\} and \{$4,1,2$\} are responsible for point attractors. Here, every attractor is associated with some confused configurations, and some initial configurations with deterministic destination. This system can catch the complex vote sharing political alliance dynamics of Indian politics where large number of regional political parties under federal structure play role in the dynamics of vote sharing. Under this CA system, a confused configuration is not associated with every attractors. Therefore, an alliance of some attractors may cover (or optimize) all possible destination of a confused configuration. Similarly, in Indian politics, the formation of alliance by regional parties target to cover the swing voters based on language, religion, caste etc.  

\begin{figure}
\centering 
\begin{tikzpicture}[scale=0.65]
\begin{axis}[
    title={},
   ylabel={$P_{104}^{\textbf{011}}$($\textbf{0}$)},
    xlabel={Number of 011 in homogeneous initial configuration \textbf{011}},
    xmin=6, xmax=24,
    ymin=0, ymax=1,
    xtick={6,9,12,15,18,21,24},
    ytick={0.00,0.20,0.40,0.60,0.80,1.00},
    legend pos=north west,
    ymajorgrids=true,
    grid style=dashed,
]
 
\addplot[
    color=blue,
    mark=square,
    ]
    coordinates {
    (6,1.00)(9,0.00)(12,0.72)(15,0.00)(18,0.44)(21,0.00)(24,0.27)
    };
    \legend{\textbf{011}}
 
\end{axis}
\end{tikzpicture}
\caption{\scriptsize{ $P_{104}^{\textbf{011}}$($\textbf{0}$) as a function of number of $011$ in the homogeneous initial configuration \textbf{011} for ACA 104.}} \label{plot1}
\end{figure}
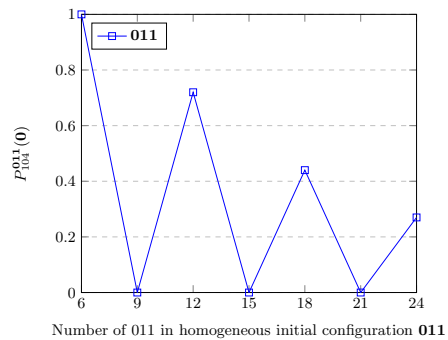

We should also mention the surprising behaviour of multiple attractor partially eccentric cloud ACA $104$ where the size of the CA matters. Here, passive primary RMT sets \{$0$\} and \{$4,1,3,6$\} are responsible for point attractors (see Table~\ref{Trules}). For ACA $104$, the probability of convergence to point attractor \textbf{0} starting from confused configurations depends on the CA size. As an evidence, Fig.~\ref{plot1} shows the $P_{104}^{\textbf{011}}$($\textbf{0}$) starting from homogeneous confused configuration \textbf{011} where $P_{104}^{\textbf{011}}$($\textbf{0}$) = $0$ for odd CA size. In details, during the evolution of the homogeneous confused configuration \textbf{011}, an intermediate configuration can be viewed as a $1$-block (i.e. consecutive cells in state $1$) which is surrounded by cell(s) in state $0$, see example configurations `$0111111$' (even size $1$-block) and `$011111$' (odd size $1$-block). Active RMTs $2$ and $5$ play the reason behind this. The size of the $1$-block is either even or odd depending on the CA size. Hereafter, the $1$-block is separated into two parts by a cell in state $0$ as RMT $7$ is active. As an example, one of the possible transition starting from even size $1$-block is $0111111 \rightarrow  0110111$. Firstly, an even size $1$-block always produces one even size $1$-block which finally converges to the smallest (two) even size $1$-block (i.e. \{$4,1,3,6$\}). Therefore, point attractor \textbf{0} is not reachable starting from even size $1$-block depending on the CA size. However, an odd size $1$-block may always produce two odd size $1$-block which may finally converges to point attractor \textbf{0}. As an example, one of the possible transition starting from odd size $1$-block is $011111 \rightarrow 010111$.


Apart from the dual property, we have also identified multiple attractor eccentric, partially eccentric, deterministic cloud system by restricting confused journey corresponding to every set of point attractors. As an example, for point attractors with primary RMT sets \{$0$\} and \{$4,1,3,6$\}, the restriction condition to avoid confused journey is active RMT $2$ and passive RMT $5$. Here, active RMT $2$ is accountable to reach the point attractor. As an example, one of the possible transition is $010110$ $\langle 1\textbf{2}5364 \rangle$ $\rightarrow$ $000110$ $\langle 001364 \rangle$. Whereas active RMT $5$ may responsible for confusion, see example transition $010110$ $\langle 12\textbf{5}364 \rangle$ $\rightarrow$ $011110$ $\langle 137764 \rangle$. For other set of point attractors, it can be shown by similar logic. Here, we have identified ACAs \textbf{4},~\textbf{36},~\textbf{132},~\textbf{164} (resp. \textbf{12},~\textbf{44},~\textbf{72},~\textbf{104},~\textbf{172},~\textbf{232}) as eccentric (resp. partially eccentric) cloud system. Whereas, ACAs \textbf{140},~\textbf{200},~\textbf{204} are deterministic cloud system. 

To sum up, we have already classified two-attractor and multiple attractor ACAs of Table~\ref{t32} into eccentric, partially eccentric and deterministic cloud system. Finally, we state our main theorem for $32$ convergent ACAs of Table~\ref{t32}.

 \begin{theoremm}
 \label{main}
 (Main result) Under asynchronous updating scheme, among the $32$ minimal representing converging ACAs with multiple point attractors, $8$,~$14$ and $10$ rules are eccentric, partially eccentric, deterministic cloud ACAs respectively.
 
\begin{table}[!htbp] 
\centering
\scriptsize
\begin{tabular}{ll} \hline  
Eccentric cloud ACA & \textbf{4,36,132,162,164,170,178,184}  \\
Partially eccentric cloud ACA & \textbf{5,12,13,44,72,76,77,78,94,104,} \\ 
 &\textbf{160,168,172,232} \\
Deterministic cloud ACA & \textbf{128,130,136,138,140,146,152,}\\
 & \textbf{154,200,204}\\
\hline
\end{tabular}
\end{table}
\end{theoremm}

\bibliography{ref}

\end{document}